# Hyper-honeycomb iridate $\beta$-Li$_2$IrO$_3$ as a platform for Kitaev magnetism

Tomohiro Takayama[1,2*], Akihiko Kato[2], Robert Dinnebier[1], Jürgen Nuss[1] and Hidenori Takagi[1,2]

[1]Max Planck Institute for Solid State Research, Heisenbergstrasse 1, 70569 Stuttgart, Germany
[2]Department of Physics and Department of Advanced Materials, University of Tokyo, 7-3-1 Hongo, Tokyo 113-0033, Japan

## Abstract

The realization of Kitaev spin liquid, where spins on a honeycomb lattice are coupled ferromagnetically by bond-dependent anisotropic interactions, has been a sought-after dream. 5$d$ iridium oxides $\alpha$-Li$_2$IrO$_3$ and $\alpha$-Na$_2$IrO$_3$ with a honeycomb lattice of $J_{eff}$ = 1/2 moments recently emerged as a possible materialization. Strong signature of Kitaev physics, however, was not captured. Here we report the discovery of a complex iridium oxide $\beta$-Li$_2$IrO$_3$ with $J_{eff}$ = 1/2 moments on "hyper-honeycomb" lattice, a three-dimensional analogue of honeycomb lattice. A positive Curie-Weiss temperature $\theta_{CW}$ ~ 40 K indicated dominant ferromagnetic interactions among $J_{eff}$ = 1/2 moments in $\beta$-Li$_2$IrO$_3$. A magnetic ordering with a small entropy change was observed at $T_c$ = 38 K, which, with the application of magnetic field of only 3 T, changed to a fully polarized state of $J_{eff}$ = 1/2 moments. Those results imply that hyper-honeycomb $\beta$-Li$_2$IrO$_3$ is located in the vicinity to a Kitaev spin liquid.

The realization of a quantum spin-liquid has been one of the central issues in the field of condensed matter physics, and geometrically frustrated lattice comprising a triangular motif is at the heart of the exploration[1]. When spins interact antiferromagnetically on a triangle-based lattice, they cannot simultaneously satisfy every magnetic coupling and macroscopic degeneracy remains in the ground state. The promising candidates include organic salts with a triangular lattice[2], transition-metal kagome oxides[3] and the hyper-kagome iridium oxide[4]. Another route to a gapless spin liquid was proposed by Kitaev, which is based on bond-dependent anisotropic coupling among spins on a honeycomb lattice[5]. On the honeycomb lattice, each site has three neighbors connected by 120° bonds. If only *x*, *y* and *z* components of spins couple ferromagnetically along the three bonds respectively, the bond-dependent polarization of spins conflicts with each other, leading to a frustration. The Kitaev model with such bond-frustration is exactly solvable. The ground state is known to be a quantum spin liquid, which may be viewed as a RVB state of ferromagnetic spin pairs on the bonds. It has attracted interest not only because of the exotic quantum liquid state but also the connection to quantum computing. The realization of Kitaev model in a crystalline solid, however, has been quite elusive. Recently, layered honeycomb oxides, α-$Li_2IrO_3$ and α-$Na_2IrO_3$ were proposed to substantiate Kitaev physics[6-8].

In the layered honeycomb iridates with $Ir^{4+}$ ions octahedrally coordinated with $O^{2-}$ ions, the spin-orbital entangled $J_{eff}$ = 1/2 states[9] could give rise to the bond-dependent anisotropic coupling through an interference of superexchange paths. $Ir^{4+}$ ions accommodate five 5*d* electrons. The large splitting between the $t_{2g}$ and $e_g$ manifolds, due to the octahedral crystalline field, allocates all the five electrons into the $t_{2g}$ manifold. The spin-orbit coupling of heavy Ir, as large as 0.5 eV, reconstructs the $t_{2g}$ manifold into lower $J_{eff}$ = 3/2 quartet with four electrons and upper $J_{eff}$ = 1/2 doublet with one electron. Localized $J_{eff}$ = 1/2 moments are produced by the presence of a modest on-site Coulomb $U$ in the half-filled $J_{eff}$ = 1/2 band. The $J_{eff}$ = 1/2 state consists of equal superposition of the three $t_{2g}$ orbitals with real and imaginary orbital components and opposite spins, $|J_{eff} = 1/2> = \frac{1}{\sqrt{3}}(|xy, \pm\sigma> \pm |yz, \mp\sigma> + i|zx, \mp\sigma>$). In α-$Li_2IrO_3$ and α-$Na_2IrO_3$, $IrO_6$ octahedra are connected by sharing their edges. The neighboring $J_{eff}$ = 1/2 pseudospins interact with each other through the *p-d* hybridization associated with the planer Ir-$O_2$-Ir bond. If the two Ir-O-Ir bonds are in the ideal 90° configuration, the existence of imaginary orbital component gives rise to a destructive interference between the two Ir-O-Ir superexchange paths[6]. This results in a bond-dependent ferromagnetic interaction, identical to those in the Kitaev model.

Emergent Kitaev physics in the honeycomb iridates has triggered intensive investigations both experimentally and theoretically. $\alpha$-Li$_2$IrO$_3$ and $\alpha$-Na$_2$IrO$_3$ order antiferromagnetically at a low temperature around 15 K[8,10]. The Curie Weiss temperatures estimated from the magnetic susceptibility $\chi(T)$ at high temperatures are negative, ~-125 K and ~-40 K for Na$_2$IrO$_3$ and Li$_2$IrO$_3$, respectively. This means that antiferromagnetic interactions, stronger than the ferromagnetic superexchange coupling, are present[11]. The magnetic ordering of Na$_2$IrO$_3$ was found to be a zig-zag type[12,13] which could be ascribed to the coexistence of antiferromagnetic interactions with Kitaev ferromagnetic coupling[14,15] but the role of Kitaev physics seems secondary. There exists a distance to the Kitaev limit of the gapless spin-liquid state in $\alpha$-Na$_2$IrO$_3$ and $\alpha$-Li$_2$IrO$_3$.

The two honeycomb iridates, $\alpha$-Na$_2$IrO$_3$ and $\alpha$-Li$_2$IrO$_3$, have been the sole playground for Kitaev physics to date. Other iridates accommodating edge-sharing IrO$_6$ network show only strong antiferromagnetic coupling[4] or metallic behavior[16]. In search for a new platform for Kitaev physics, we discovered a new form of Li$_2$IrO$_3$, $\beta$-Li$_2$IrO$_3$, consisting of a three-dimensional analogue of honeycomb lattice which we call "hyper-honeycomb" lattice. The $\chi(T)$ of hyper-honeycomb $\beta$-Li$_2$IrO$_3$ showed the presence of dominant ferromagnetic coupling, indicative of the predominant Kitaev-type interaction. A magnetic ordering, likely non-collinear, was observed at 38 K, which turns into a fully polarized state of $J_{eff} = 1/2$ pseudospins under magnetic fields above 3 T. Those results, together with the theoretical studies on extended Kitaev model for hyper-honeycomb lattice[17], place $\beta$-Li$_2$IrO$_3$ in a critical proximity to the three-dimensional Kitaev spin liquid.

**Results**

**Crystal structure analysis.** The polycrystalline samples of $\beta$-Li$_2$IrO$_3$ were synthesized by a solid state reaction from Li$_2$CO$_3$, IrO$_2$ and LiCl (see Methods). The obtained powder was found to consist of a new phase and a small trace of IrO$_2$. Any trace of the layered honeycomb $\alpha$-Li$_2$IrO$_3$ was not observed. The powder X-ray diffraction pattern is displayed in Fig. 1**a**. The detailed structure was then refined by single crystal X-ray analysis using 50 μm-size crystal grains. The result of refinement is summarized in Table 1. The lithium content with respect to Ir was found to be ~2 within our resolution by ICP (Inductively Coupled Plasma) analysis. Both elemental analysis and refinement of the occupation factors of the lithium sites during structure determination gave no hint for a defect in the lithium sub-lattice.

The crystal structure of $\beta$-Li$_2$IrO$_3$ thus determined is illustrated in Fig. 1**b-c**, which is based on the rock salt type of structure. It can be described as a distorted cubic closed packed arrangement of oxygen atoms with iridium and lithium atoms occupying all octahedral holes in a specific ordered manner (Fig.

1**c**). $\beta$-Li$_2$IrO$_3$ is isostructural to $\beta$-Na$_2$PtO$_3$[18] and Ba$_3$SiI$_2$[19], latter one with inverted anion and cation distribution. The local structure around an iridium atom is closely related to that of $\alpha$-Li$_2$IrO$_3$. Each IrO$_6$ octahedron is connected with three neighboring IrO$_6$ octahedra by sharing its three edges (Fig. 1**d**), which gives rise to three Ir-O$_2$-Ir planar bonds with the planes almost perpendicular to each other as in $\alpha$-Li$_2$IrO$_3$. If $J_{eff}$ = 1/2 moment were placed on the Ir site, the exchange interaction via Ir-O$_2$-Ir should give rise to the competing $x$, $y$ and $z$ spin polarizing bonds as in $\alpha$-Li$_2$IrO$_3$.

The network of iridium ions in $\beta$-Li$_2$IrO$_3$, depicted in Fig. 2**a**, is topologically linked to a honeycomb lattice. Let us consider a stack of 2D honeycomb lattices. The 2D honeycomb lattice can be viewed as planar zig-zag chains (colored in pink) connected at the corners with bridging bonds (black dotted line), as seen in Fig. 2**b**. In the sub-lattice of $\beta$-Li$_2$IrO$_3$, the zig-zag Ir chains are connected by the bridging bonds parallel the $c$-axis as in the 2D honeycomb. In contrast to the 2D honeycomb lattice, however, the zig-zag chains are alternately rotated by 69.9° around the $c$-axis (pink and blue chains in Fig. 2**a**) and connected to the zig-zag chains in the layers above and below. We emphasize again here that the local environment of each iridium atom is identical with that of 2D honeycomb lattice with three 120° bonds. Because of the close link to a honeycomb structure, the Ir sub-lattice in $\beta$-Li$_2$IrO$_3$ may be called as "hyper-honeycomb". In the hyper-honeycomb Ir lattice, all the angles between the three Ir-Ir bonds are very close to 120°, and the distances between Ir atoms are almost equivalent (only ~0.1% difference). The hyper-honeycomb lattice of iridium atoms has the same topology as the silicon network in $\alpha$-ThSi$_2$, but with different deltahedral angle between two three-connected iridium atoms, 69.9° ($\beta$-Li$_2$IrO$_3$) versus 90° (ThSi$_2$). ThSi$_2$ itself, can also exist in two modifications, hyper-honeycomb ($\alpha$-ThSi$_2$) and 2D honeycomb ($\beta$-ThSi$_2$)[20], where the former consists of 10-site loops of Si atoms in contrast to the 6-site ones in the latter.

As an extension of the Kitaev model, the lattice equivalent to the hyper-honeycomb lattice, with competing $x$, $y$ and $z$ bonds, was studied theoretically[17]. The model can be mapped onto the Kitaev model and is exactly solvable. The ground state was found to be a spin-liquid state as in the original Kitaev model. We may therefore anticipate Kitaev physics and a possible spin-liquid state in $\beta$-Li$_2$IrO$_3$.

**Electric and Magnetic Properties.** Resistivity measurement indicated that $\beta$-Li$_2$IrO$_3$ is an insulator. Combined with the presence of localized moments described below, we conclude that $\beta$-Li$_2$IrO$_3$ is a spin-orbital Mott insulator with $J_{eff}$ = 1/2 moment as in $\alpha$-Li$_2$IrO$_3$[10]. The temperature dependence of magnetic susceptibility $\chi(T)$, measured on the polycrystalline sample, is shown in Fig. 3**a**. The

Curie-Weiss fitting at high temperatures between 200 K and 350 K yielded an effective moment 1.61 $\mu_B$, close to that of $J_{eff}= 1/2$ moment, and a positive Curie-Weiss temperature $\theta_{CW} \sim 40$ K. This indicates a dominant ferromagnetic interaction presumably associated with Ir-$O_2$-Ir super-exchange interaction. The bond-dependent ferromagnetic interaction is the key ingredient of the (extended) Kitaev model. With decreasing temperature, $\chi(T)$ showed a steep increase at low temperatures below ~50 K, followed by a sharp cusp at $T_c = 38$ K indicative of a magnetic ordering. The specific heat $C(T)$ showed an anomaly at $T_c = 38$ K, evidencing a second order magnetic phase transition. $\chi(T)$ did not show a decrease below $T_c$, in contrast to those of collinear antiferromagnets. It is therefore likely that the ground state is not a simple collinear antiferromagnet.

The ground state is in fact very close to ferromagnetism. The magnetization curve at 5 K (Fig. 4) clearly showed a magnetic-field induced change to a ferromagnetic state. At low fields, the magnetization increased linearly with field. With further increasing magnetic field, a kink was observed at $\mu_0 H_c \sim 3$ T, followed by a saturation behavior above $\mu_0 H_c \sim 3$ T. Any indication of meta-magnetism with a discontinuous jump was not observed in the magnetization curve down to 5 K. The magnitude of magnetization above $\mu_0 H_c \sim 3$ T was remarkably large, ~0.35 $\mu_B$/Ir. The ordered moment in other iridates such as $Sr_2IrO_4$ and $\alpha$-$Na_2IrO_3$ was reported to be around 0.20-0.36 $\mu_B$/Ir[21,22] and 0.22 $\mu_B$/Ir[13], respectively. The high field ferromagnetic state in the hyper-honeycomb iridate therefore should be essentially a fully polarized state of $J_{eff} = 1/2$ moment, which is in marked contrast to the weak ferromagnetism arising from the canted moments in $Sr_2IrO_4$[9]. The fact that almost fully polarized state is realized in the polycrystalline sample consisting of randomly oriented grains at a relatively low field of 3 T means a small anisotropy and hence very likely almost degenerate $x$, $y$ and $z$ spin polarized states.

Under a magnetic field of $\mu_0 H = 4$ T, the cusp seen in the low field $\chi(T)$ faded out as shown in the inset of Fig. 4. In accord with this, the peak in $C/T$ was smeared out, consistent with ferromagnetic ordering of moments. The magnetic entropy associated with the transition, estimated from the specific heat anomaly, is at most a few percent of $R\ln 2$, which supports for the presence strong fluctuations and hence the low lying excitations.

**Discussions**

The close proximity to the fully polarized $J_{eff} = 1/2$ state and the presence of low lying excitations point that the hyper-honeycomb $\beta$-$Li_2IrO_3$ is located much closer vicinity to a Kitaev liquid than the layered honeycomb $\alpha$-Na$_2$IrO$_3$ and $\alpha$-Li$_2$IrO$_3$. Though very close to the Kitaev spin liquid state, however, the bond dependent ferromagnetic interactions can be still superimposed by other antiferromagnetic interactions, including the deviation of Ir-O-Ir bond angle from 90° and the direct $d$-$d$ exchange, which very likely stabilize the magnetically ordered state observed in $\beta$-Li$_2$IrO$_3$. The magnetization behavior below $T_c$ = 38 K points to a non-collinear ordering such as spiral magnetic ordering. Spiral order is in fact envisaged to manifest itself at the critical boundary to the Kitaev liquid in the theoretical phase-diagrams of the 2D honeycomb iridate[10, 23].

In the 2D honeycomb $\alpha$-Li$_2$IrO$_3$ and $\alpha$-Na$_2$IrO$_3$, the weak signature of Kitaev physics was discussed at least in part due to the local distortion of IrO$_6$ octahedra[10]. The deviation of the Ir-O-Ir angle from 90°, ~95° for $\alpha$-Li$_2$IrO$_3$[7] and ~98° for $\alpha$-Na$_2$IrO$_3$[24], should suppress the ferromagnetic coupling representing Kitaev physics. The two Ir-O bonds forming 90° Ir-O-Ir bond are inequivalent in the honeycomb iridates. The lengths are as much as ~5.7% different in $\alpha$-Li$_2$IrO$_3$[7]. In hyper-honeycomb $\beta$-Li$_2$IrO$_3$, all the Ir-O-Ir angles are almost ~94.5°, closer to 90° compared with the 2D honeycomb iridates. The difference in the lengths of inequivalent Ir-O bonds are less than 0.1 %, an order of magnitude smaller that of $\alpha$-Li$_2$IrO$_3$. We argue that those local environments close to the ideal setting for the Kitaev model might be the origin of dominant ferromagnetic interaction and thus the close proximity to the spin-liquid state in $\beta$-Li$_2$IrO$_3$.

All the data presented here indicated that the hyper-honeycomb iridate $\beta$-Li$_2$IrO$_3$ is the most promising playground to challenge the realization of a Kitaev spin-liquid. One of the approaches to bring the system into the Kitaev limit could be an application of physical or chemical pressure to tune the local lattice distortion. The hyper-honeycomb iridate might offer other intriguing prospects. So far the Kitaev liquid was proposed to display unconventional superconductivity by carrier doping[25, 26], but the 2D honeycomb iridates have been yet to be tuned into a metallic state. The 3D structure of the hyper-honeycomb lattice might be advantageous in terms of bandwidth and the crystal structure free from the stacking disorder. Aside from Kitaev physics, the 2D honeycomb iridates have been also discussed to host non-trivial topological properties from the viewpoint of the weak-coupling picture[27, 28]. The hyper-honeycomb lattice, unlike the honeycomb lattice seen in the celebrated graphene, would provide a fresh fuel in search for novel topological states of matter.

**Methods**
**Sample synthesis.** Polycrystalline samples of $\beta$-Li$_2$IrO$_3$ were synthesized by a solid state reaction of

Li$_2$CO$_3$, IrO$_2$ and LiCl in a molar ratio of 10:1:100. The mixed powder was cold-pressed into a pellet and placed in an alumina crucible. The pellet was heated at 1100ºC for 24 h, cooled to 700 ºC at a rate of 30 K/h and finally furnace-cooled to room temperature.

**X-ray structural analysis and characterizations**. X-ray powder diffraction data of the powdered *β*-Li$_2$IrO$_3$ sample were collected at room temperature with a Stoe Stadi-P transmission diffractometer (primary beam Johann-type Ge (111) monochromator for Ag-*Kα*1-radiation ($\lambda$ = 0.55941 Å), Mythen-Dectris PSD with 12° 2*θ* opening) with the sample sealed in a glass capillary of 0.5 mm diameter (Hilgenberg, glass No. 50). The powder pattern was recorded for 5.5 h in the range from 2-45° 2*θ* with a step width of 0.012° 2*θ*. The sample was spun during measurement for better particle statistics. Rietveld refinement was performed using the program TOPAS Version 4.2 (Bruker AXS, 2010). The crystal structure of single crystals was analyzed with a three circle X-ray diffractometer (Bruker AXS) equipped with SMART APEX CCD, and Mo *Kα* radiation (see Supplementary for details). Electric, Magnetic and thermodynamic properties were measured by Quantum Design PPMS & MPMS.

superconductivity from the Kitaev-Heisenberg model and possible application to (Na$_2$/Li$_2$)IrO$_3$. *Phys. Rev. B* **86,** 085145 (2012).

27. Shitade, A., Katsura, H., Kunes, J., Qi, X. L., Zhang, S. C., & Nagaosa, N. Quantum spin Hall effect in a transition metal oxide Na$_2$IrO$_3$. *Phys. Rev. Lett.* **102,** 256403 (2009).

28. Kim, C. H., Kim, H. S., Jeong, H., Jin, H., & Yu, J. Topological quantum phase transition in 5*d* transition metal oxide Na$_2$IrO$_3$. *Phys. Rev. Lett.* **108,** 106401 (2012).

29. Momma, K. & Izumi, F. Vesta 3 for three-dimensional visualization of crystal, volumetric and morphology data. *J. Appl. Cryst.* **44,** 1272-1276 (2011).

**Acknowledgements**
We thank A. W. Rost for invaluable discussions and critical reading of the manuscript. We are grateful to B. J. Kim, A. Jain, D. Haskel, L. Hozoi, J. Nasu, M. Udagawa, Y. Motome and G. Jackeli for fruitful discussion. This work was partly supported by Grant-in-Aid for Scientific Research (S) (Grand No. 24224010).

**Author contributions**
T.T. conceived and designed the project. T.T. and A.K. performed synthesis and measurements of physical properties. R.D. collected and analyzed the powder X-ray diffraction data, and J.N. performed the single crystal analysis. H.T. supervised the project. T.T. and H.T. wrote the manuscript with inputs from all authors. All authors discussed and reviewed the paper.

**Competing financial interests**: The authors declare no competing financial interests.

**Figure Captions**
Figure 1. **Crystal structure of *β*-Li$_2$IrO$_3$. a.** Powder X-ray diffraction pattern and Rietveld analysis of *β*-Li$_2$IrO$_3$ polycrystalline sample. The red dots, green line and purple bars represent the observed pattern, the Rietveld fit profile and the reflection positions, respectively. The blue line shows the difference between the observed and calculated profiles. The high angle part starting at 23.0º is enlarged for clarity. The final *R* indices are $R_{wp}$ = 4.440% and $R_p$ = 3.374%. **b.** The perspective view of the crystal structure of *β*-Li$_2$IrO$_3$. Green, gray and blue spheres represent lithium, iridium and oxygen atoms, respectively. The IrO$_6$ octahedra are connected by sharing their edges in three dimensions. **c.** Projected view of the unit cell in the *bc*-plane. The lithium and iridium atoms are accommodated in the octahedral voids formed by the oxygen atoms comprising a distorted cubic closed packed arrangement. **d.** Local lattice

network of $IrO_6$ octahedra in $\beta$-$Li_2IrO_3$[29], displaying Ir-O bond lengths and two different Ir-O-Ir angles obtained from the single crystal analysis.

Figure 2. **Hyper-honeycomb network of iridium atoms. a.** Hyper-honeycomb lattice in $\beta$-$Li_2IrO_3$[29]. The pink and blue lines show the twisted zig-zag chains alternating along the *c*-axis. The black dotted lines are the bond bridging the zig-zag chains. The numbers indicated are Ir-Ir distances and the angles between Ir atoms. **b.** 2D honeycomb lattice shown as a reference. The zig-zag chains, which run in three directions rotated by 120º, are confined in the same plane.

Figure 3. **Magnetic transition in $\beta$-$Li_2IrO_3$. a.** Temperature dependence of magnetic susceptibility under 1 T. The inset shows the temperature dependence of inverse of magnetic susceptibility. The solid line delineates the Curie-Weiss fit at high temperatures between 200 and 350 K. **b.** Temperature dependence of specific heat divided by temperature recorded at 0 and 7 T.

Figure 4. **Magnetization curve of $\beta$-$Li_2IrO_3$.** The blue and red dots are data taken at 5K and 50 K, respectively. The inset shows the temperature dependence of magnetization under a magnetic field of 4 T.

Table 1 **Structural parameters of $\beta$-Li$_2$IrO$_3$.** The space group is *Fddd* (No. 70) and $Z = 16$, and the lattice constant is $a = 5.9194(3)$ Å, $b = 8.4562(4)$ Å and $c = 17.8271(9)$ Å. $g$ and $U_{iso}$ denote site occupancy and isotropic displacement parameter, respectively. The final $R$ indices are $R = 0.027$ and $wR = 0.0480$.

| Atom | site | g | x | y | z | $U_{iso}$ (Å$^2$) |
|---|---|---|---|---|---|---|
| Ir | 16*g* | 1 | 1/8 | 1/8 | 0.70854(2) | 0.00560(4) |
| O(1) | 16*e* | 1 | 0.8572(5) | 1/8 | 1/8 | 0.0078(4) |
| O(2) | 32*h* | 1 | 0.6311(5) | 0.3642(3) | 0.0383(1) | 0.0094(3) |
| Li(1) | 16*g* | 1 | 1/8 | 1/8 | 0.0498(5) | 0.0051(11) |
| Li(2) | 16*g* | 1 | 1/8 | 1/8 | 0.8695(7) | 0.0155(18) |

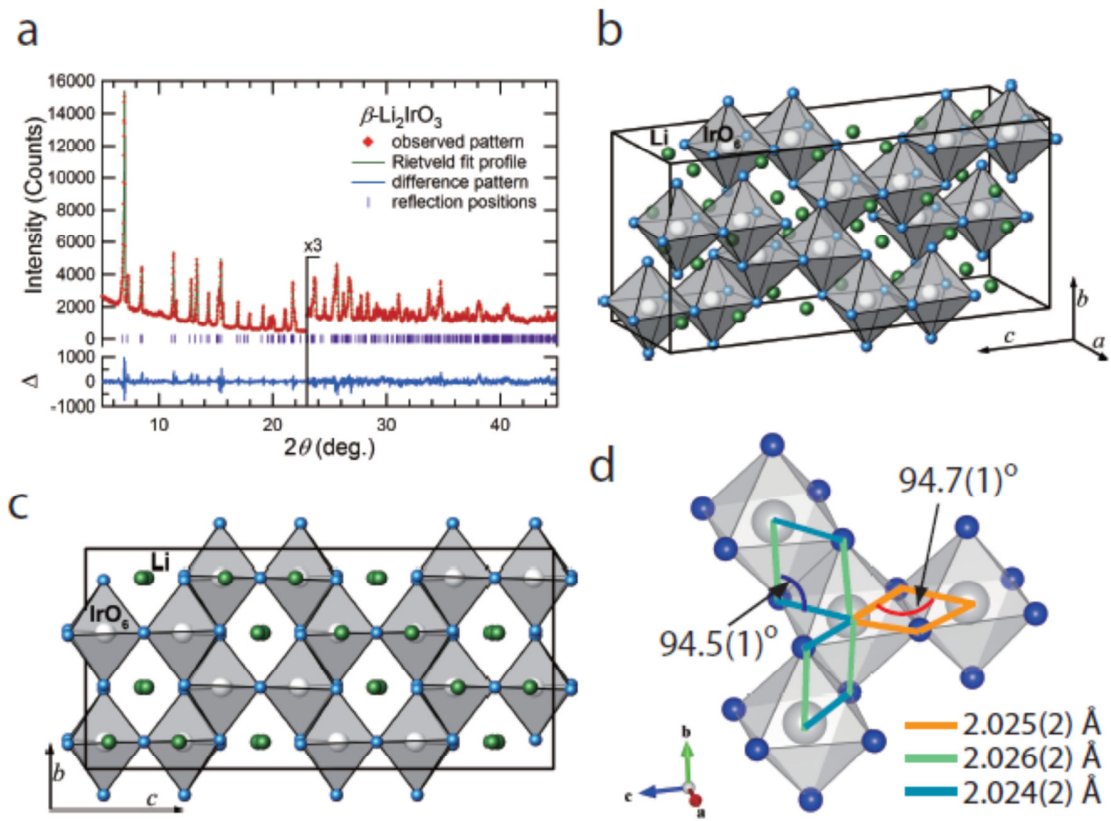

Figure 1 (T. Takayama et al.)

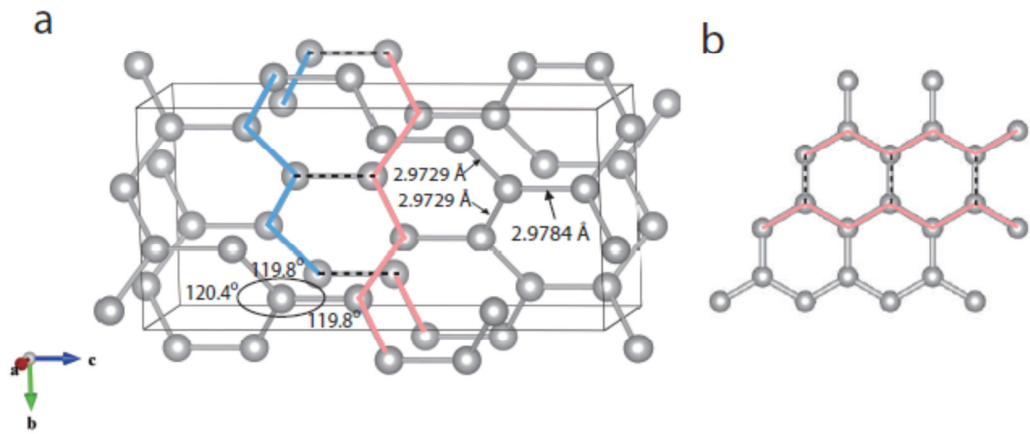

**Figure 2 (T. Takayama et al.)**

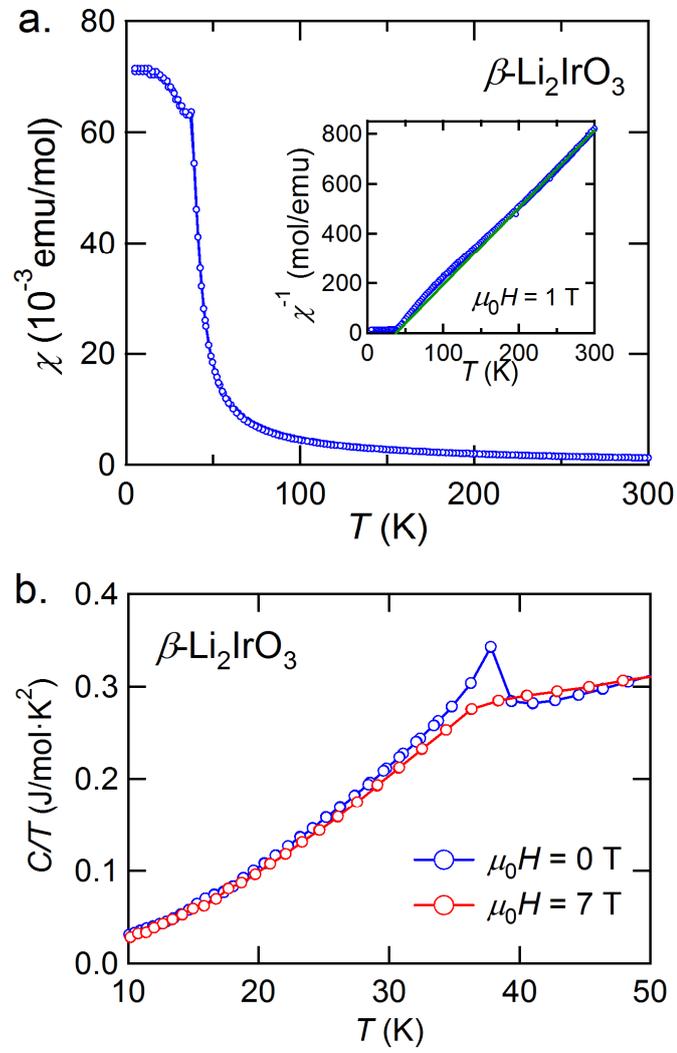

**Figure 3 (T. Takayama et al.)**

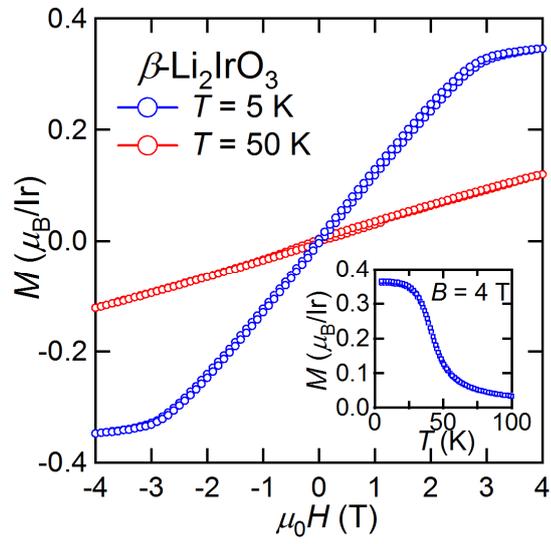

**Figure 4 (T. Takayama et al.)**